\documentclass[12pt]{article}

\title{{\small\hfill IMSc/2007/10/13}\\
\textbf{Are neutrino masses dynamically generated by quantum gravity attractions?}}
\author{
H. S. Sharatchandra$^{a}$\footnote{E-mail: sharat@imsc.res.in} \\\\
$^a$ The Institute of Mathematical Sciences, C.I.T. Campus, Taramani P.O.,\\
Chennai 600113, India}
\date{}
\begin{document}
\maketitle
\begin{abstract}
It is argued that the quantum gravity attractions dynamically generate tiny degenerate Majorana masses
for the neutrinos. The unequal masses of the charged leptons then 
induce a computable neutrino mass matrix with splittings and 
mixings through the electroweak interactions.  In this way the Standard Model
including quantum gravity can accommodate and predict the neutrino masses and mixings. Some
consequences are pointed out.
\end{abstract}
\newpage
Consider a system of massless (Dirac or Majorana) fermion and
(anti-)fermion. When the two have tiny momenta, the net kinetic
energy of the system is also tiny. Therefore a small 
attraction between the two could lead to a tachyonic ( i.e.
$(invariant \ mass)^2 <0$ ) bound state heralding an instability
of the
system. This argument is rather naive. A bound state also
involves large momenta of the constituents and subsequent
effects are not clear. Also it is not clear as to how much
and what kind of attraction suffices. In the non-relativistic case,
relevant for the BCS theory of superconductivity, a quantum
mechanical calculation  was sufficient to check 
for an instability.
But in our case we need a detailed field theoretic 
calculation. The framework now required is the Bethe-
Salpeter equation. As massless particles are involved, the
infrared divergences may generate anomalous dimensions ( in
the infrared) for the full fermion propagators and also for the
kernel of the scattering amplitude. A fairly
general analysis of the situation was made in Ref. \cite{jms}.
This was made possible by mapping the Bethe-Salpeter
equation to a generalized quantum mechanical Hamiltonian and 
by a variational estimation of the ground state energy of
this Hamiltonian. The outcome is that for a long range 
attraction
there is an instability for an arbitrarily weak coupling.
An exception is when the anomalous dimensions are such that 
the quantum mechanical Hamiltonian has a scale invariance. In
this case the instability is  at a critical value of the 
coupling constant. ( Also in case of a short range attraction,
the instability requires a large enough coupling constant.)

The instability in the two body system leads to a rearrangement
of the field theoretic vacuum resulting in a dynamical breaking  of
chiral ( or other) symmetries. The calculation of the resulting order parameter ( viz. mass) is more difficult than the variational calculation of \cite{jms}. It requires the solution 
of a `gap equation'. We need a non-perturbative solution of 
the Schwinger- Dyson equation for the fermion propagator
\cite{sd}.

If we
initially have $n$ massless fermions, the inverse of the free propagator
is $ \delta_{ab} \gamma .p$ where $a,b=1,2 \cdots n$. On the other hand,
allowing 
for the dynamical breaking of chiral symmetry,
the inverse of the full propagator has the form
 $A_{ab}(p) \gamma .p+B_{ab}(p)$ involving matrices $A$ and $B$.
If the Schwinger- Dyson equation has a 
solution with $B_{ab}(p)$  not identically zero, the chiral symmetry is dynamically broken. 
 
The Schwinger- Dyson equation for the propagator is only the
first of an infinite set of coupled non-linear equations for
the Green's functions. In practice we can only hope to solve
it with some physically motivated approximation, especially 
for the behavior of the vertices etc.An example is the
rainbow  approximation for solving the `gap equation'.
This  approximation directly corresponds
to the ladder approximation for the check of instability.
Even with these approximations we need to solve non-linear
equations \cite{sd} for $A_{ab}(p)$ and $B_{ab}(p)$.
The value at zero momentum $p=0$, of $B_{ab}(p)$  may be roughly regarded as the (dynamically generated) mass matrix 
$M_{ab}$ (instead of the zeroes of the inverse
propagator). 

It is of  great importance to apply these techniques to the 
case of long range attractions induced by quantum gravity. It is well known 
\cite{w} that the Einstein gravity, when quantized, is 
infrared free. This means that there is no anomalous dimension 
for the fermion (in the infrared). Also the interactions are 
governed by the dimensionful Newton's constant. Hence 
the generalized Hamiltonian \cite{jms} related to the Bethe-Salpeter equation  will not have scale
invariance. Therefore we are tempted to conclude that
there is an instability for any value of
Newton's constant. However the calculations of \cite{jms}
are not directly applicable to the quantum gravity case. The reason
is the equivalence principle due to which the graviton couples 
to the energy-momentum tensor. As the fermions are massless, the
coupling vanishes linearly with the energy-momentum of the 
fermion. The calculations of \cite{jms} are more relevant for spin
zero or spin one field quanta, where the coupling is finite
even at vanishing momenta of the fermions.

The check of instability in the quantum gravity case is technically
challenging. We are also hampered by the non-renormalizability
of the Einstein gravity. A possibility is to do a calculation
using a cut-off. It is interesting to see whether string
theory techniques predict  dynamical generation of masses.
In view of these issues we do not attempt a calculation here. 
We only argue for the results that we  expect. Note that
in a theory such as QCD, the scale of dynamically generated (i.e. constituent quark) masses is also the scale of the theory viz. $\Lambda_{QCD}$. Such a
result for quantum gravity is unpleasant ( because of Planck scale masses).
But we have pointed out  that  in the case of quantum gravity the coupling 
vanishes at low momenta and does not bind the fermion as much.
This could simply mean that there is no instability  at all.
We propose that in (the correct theory of ) quantum gravity,
there is a very weak instability, quantitatively very different
from the QCD case. The scale for quantum gravity, i.e. the Planck mass,
refers to the momenta at which gravity becomes strong. We have
argued that the instability involves fermions of very low momenta
with very weak attractions. Therefore we may expect that
the dynamically  generated masses are much less than the Planck scale. Let 
us presume that the solution of the 'gap equation' gives for 
the dynamically generated mass $m$, 
$ \frac{1}{8\pi^2} ln \frac{m_P}{m} \approx 1$ in terms of 
the Planck
mass $m_P$. This gives masses of $O(10^{-1}) \ eV$.

If the charged leptons and quarks already have much larger masses
through the Higgs mechanism,  then quantum gravity will not 
cause any instability in that sector and does not have 
any drastic effect on  their masses. On the other hand,
there is a dramatic effect on the massless neutrinos of the Standard
Model. Note that the gravitons induce an universal attraction
between the massless fermions of any species and also between a
fermion and an anti-fermion. Though quantum gravity attractions are 
there between a neutrino and an anti- neutrino, this system in the 
center of mass frame is necessarily in spin one channel.
It is only the pair of  neutrinos with opposite momenta and 
hence helicities that forms a spin zero system. Thus it is
natural to expect  dynamically generated  masses given by 
$ m \ \epsilon _{\alpha \beta} \sum_a \nu_{\alpha}^a 
\nu_{\beta}^a$  where $a = 1,2,3$ labels the neutrino 
flavors and $ \alpha, \beta =1,2$ are the (Weyl) spinor indices.
This means degenerate Majorana masses for the neutrinos. This is the situation when only
quantum gravity attractions are taken into account.

When we bring in the electroweak interactions, the effective interactions between
neutrinos are different for differing flavors. 
This is because of 
different masses of the charged lepton propagators in  
the box diagrams involving an exchange of a pair of charged
vector bosons $W$. This has an effect on the 
 mass matrix. We
have to solve the Schwinger- Dyson equations allowing for most
general mixing  $B_{ab}(p)$. With this mixing of different neutrino 
flavors, diagrams involving only 
$W$ exchanges without any gravitons also are involved in the equation. Thus even though raison d'\^{e}tre for the non-zero 
mass is the
quantum gravity attractions, the mass matrix gets contributions
from self energy diagrams with purely electroweak exchanges.
The reason is that the equation for $B_{ab}(p)$ is non-linear.
This gives splittings and mixings in
the neutrino mass matrix which are not negligible.
We can state this in a different way.
We have argued that it is the scattering amplitudes at
low momenta which are relevant for the dynamically generated tiny masses. 
The contribution of quantum gravity to this amplitude is small due to
vanishing couplings at low momenta. But the contributions 
due to electroweak interactions are finite and flavor dependent even in the limit of zero momenta. 
Thus the  electroweak interactions give  
splittings and mixings that are quite large.

A major consequence of our proposal is the Majoron \cite{mj}, the Goldstone 
boson of the broken (B-L) symmetry. ( The condensate breaks the 
lepton number L spontaneously, but the $U(1)_A$ anomaly of the
Standard Model means that ‎it is strictly the (B-L) number which
is spontaneously broken.) There has been an extensive analysis
\cite{cal} of the phenomenological consequences of the  Majoron for Cosmology, 
Astrophysics and also lab experiments.These are applicable for 
the present proposal also.

We have argued that the instability caused by quantum gravity attractions is 
very weak due to the vanishing couplings to the massless fermions 
at low momenta. This scenario also means that the condensate and the dynamically generated masses are very fragile in hot  and dense environments.
Whereas the condensate is entirely due to quantum gravity, the splittings
and mixings are due to electroweak interactions and are quite robust. This points to an
unusual behavior of the neutrinos in Cosmology and Astrophysics.

\end{document}